# Reliable Multi-Path Routing Schemes for Real-Time Streaming

Emin Gabrielyan
*Switzernet Sàrl and EPFL*
*Lausanne, Switzerland*
emin.gabrielyan@switzernet.com

Roger D. Hersch
*École Polytechnique Fédérale*
*de Lausanne (EPFL), Switzerland*
rd.hersch@epfl.ch

## Abstract

*In off-line streaming, packet level erasure resilient Forward Error Correction (FEC) codes rely on the unrestricted buffering time at the receiver. In real-time streaming, the extremely short playback buffering time makes FEC inefficient for protecting a single path communication against long link failures. It has been shown that one alternative path added to a single path route makes packet level FEC applicable even when the buffering time is limited. Further path diversity, however, increases the number of underlying links increasing the total link failure rate, requiring from the sender possibly more FEC packets. We introduce a scalar coefficient for rating a multi-path routing topology of any complexity. It is called Redundancy Overall Requirement (ROR) and is proportional to the total number of adaptive FEC packets required for protection of the communication. With the capillary routing algorithm, introduced in this paper we build thousands of multi-path routing patterns. By computing their ROR coefficients, we show that contrary to the expectations the overall requirement in FEC codes is reduced when the further diversity of dual-path routing is achieved by the capillary routing algorithm.*

## 1. Introduction

Packetized IP communication behaves like an erasure channel. Information is chopped into packets, and each packet is either received without error or not received. Packet level erasure resilient FEC codes can mitigate packet losses by adding redundant packets, usually of the same size as the source packets.

In off-line streaming erasure resilient codes achieve extremely high reliability in many challenging network conditions [MacKay05]. For example, it is possible to deliver voluminous files (e.g. recurrent updates of GPS maps) via satellite broadcast channel without feed-backs to millions of motor vehicles under conditions of fragmental visibility (see [Honda04] and Raptor codes [Shokrollahi04]). In the film industry, the day's film footage can be delivered from the location it has been shot to the studio that is many thousands of miles away not via FedEx or DHL, but over the lossy internet even with long propagation delays (see [Hollywood03] and LT codes [Luby02]). Third Generation Partnership Project (3GPP), recently adopted Raptor [Shokrollahi04] as a mandatory code in Multimedia Broadcast/Multicast Service (MBMS). The benefit of off-line streaming from application of FEC relies on time diversity, i.e. on the receiver's right to not forward immediately to the user the received information. Long buffering is not a concern, the receiver can unrestrictedly hold the received packets, and as a result packets representing the same information can be collected at very distant periods of time.

In real-time single-path streaming FEC can only mitigate short failures of fine granularity. See [Choi06] using RS(24,20) packet level code with 20 source packets and 4 redundant packets or also [Johansson02], [Huang05], [Padhye00] and [Altman01]. Due to restricted playback buffering time, packets representing the same information cannot be collected at very distant periods of time. Instead of relying on time-diversity FEC in real-time streaming can rely on path-diversity. Recent publications show the applicability of FEC in real-time streaming with dual-path routes. Author of [Qu04] shows that strong FEC sensibly improves video communication following two disjoint paths and that in two correlated paths weak FEC codes are still advantageous. [Tawan04] proposes adaptive multi-path routing for Mobile Ad-Hoc Networks (MANET) addressing the load balance and capacity issues, but mentioning also the potential advantages for FEC. Authors of [Ma03] and [Ma04] suggests replacing in MANET the link level Automatic Repeat Query (ARQ) by a link level FEC assuming regenerating nodes. Authors of [Nguyen02] and [Byers99] studied video streaming from multiple servers. The same author [Nguyen03] later studied real-time streaming over a dual-path route using a static Reed-Solomon RS(30,23) code (FEC blocks carrying 23 source packets and 7 redundant packets).

[Nguyen03], similarly to [Qu04], compares dual-path scenarios with the single OSPF routing strategy and has shown clear advantages of the dual-path routing. The path diversity in all these studies is limited to either two (possibly correlated) paths or in the most general case to a sequence of parallel and serial links. Various routing topologies have so far not been regarded as a space to search for a FEC efficient pattern.

In this paper we try to present a comparative study for various multi-path routing patterns. Single path routing is excluded from our comparisons, being considered too hostile. Steadily diversifying routing patters are built layer by layer with the *capillary routing* algorithm (sections 2).

In order to compare multi-path routing patterns, we introduce Redundancy Overall Requirement (ROR), a routing coefficient relying on the sender's transmission rate increases in response to individual link failures. By default, the sender is streaming the media with static FEC codes of a constant weak strength in order to tolerate a certain small packet loss rate. The packet loss rate is measured at the receiver and is constantly reported back to the sender with the opposite flow. The sender increases the FEC overhead whenever the packet loss rate is about to exceed the tolerable limit. This end-to-end adaptive FEC mechanism is implemented entirely on the end nodes, at the application level, and is not aware of the underlying routing scheme [Kang05], [Xu00], [Johansson02], [Huang05] and [Padhye00]. The overall number of transmitted adaptive redundant packets for protecting the communication session against link failures is proportional *(1)* to the usual packet transmission rate of the sender, *(2)* to the duration of the communication, *(3)* to the single link failure rate, *(4)* to the single link failure duration and *(5)* to the ROR coefficient of the underlying routing pattern followed by the communication flow. The novelty brought by ROR is that a routing topology of any complexity can be rated by a single scalar value (section 3).

In section 4, we present ROR coefficients of different routing layers built by the capillary routing algorithm. Network samples are obtained from a random walk MANET with several hundreds of nodes. We show that path diversity achieved by the capillary routing algorithm reduces substantially the amount of redundant FEC packets required from the sender.

## 2. Capillary routing

In subsection 2.1 we present a simple Linear Programming (LP) method for building the layers of capillary routing. A more reliable algorithm is described in subsection 2.2. In subsection 2.3 we present the discovery of bottlenecks at each layer of capillary routing, required for construction of successive layers.

### 2.1. Basic construction

Capillary routing can be constructed by an iterative LP process transforming a single-path flow into a capillary route. First minimize the maximal value of the load of all links by minimizing an upper bound value applied to all links. The full mass of the flow will be split equally across the possible parallel routes. Find the bottleneck links of the first layer (see subsection 2.3) and fix their load at the found minimum. Minimize similarly the maximal load of all remaining links without the bottleneck links of the first layer. This second iteration further refines the path diversity. Find the bottleneck links of the second layer. Minimize the maximal load of all remaining links, but now without the bottlenecks of the second layer as well. Repeat this iteration until the entire communication footprint is enclosed in the bottlenecks of the constructed layers.

Fig. 1, Fig. 2 and Fig. 3 show the first three layers of the capillary routing on a small network. The top node on the diagrams is the sender, the bottom node is the receiver and all links are oriented from top to bottom.

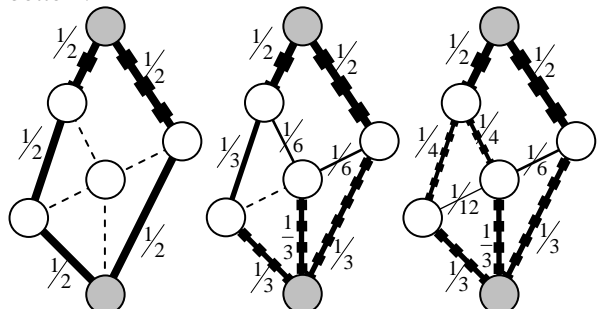

**Fig. 1.** In the first layer the flow is equally split across two paths, two links of which, marked by thick dashes, are the bottlenecks.

**Fig. 2.** The second layer minimizes to 1/3 the maximal load of the remaining seven links and identifies three bottlenecks.

**Fig. 3.** The third layer minimizes to 1/4 the maximal load of the remaining four links and identifies two bottlenecks.

Fig. 4 shows the 10-th layer of capillary routing between a pair of end nodes on a network with 180 nodes and 1374 links. Links not carrying traffic are not shown. The solid lines of the diagram represent 55 bottleneck links belonging to one of the 10 layers. The dashed lines represent a min-cost solution of the remaining flow not enclosed in bottlenecks after the 10-th layer. There could be several tens of additional routing layers until complete capillarization is achieved.

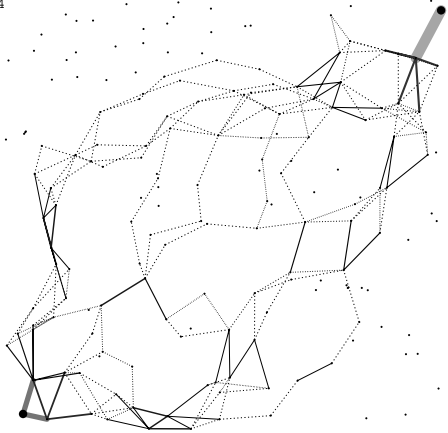

**Fig. 4.** Routing pattern of layer 10 built by the capillary routing algorithm on a network sample with 150 nodes

## 2.2. Numerically stable version

Although the described LP process is completely valid, it is numerically instable. The precision errors propagating through the layers of capillary routing reach noticeable sizes and, when dealing with tiny loads, result in infeasible LP problems. We have found a different, stable LP method which maintains the values of parameters and variables in the same order of magnitude at all times.

Instead of decreasing the maximal value of loads of the links, the routing path is discovered by solving max flow problems defined by the flow-out coefficients at each node. Initially only the peer nodes have non-zero flow-out coefficients: +1 for the source and –1 for the sink (Fig. 5 and Fig. 6).

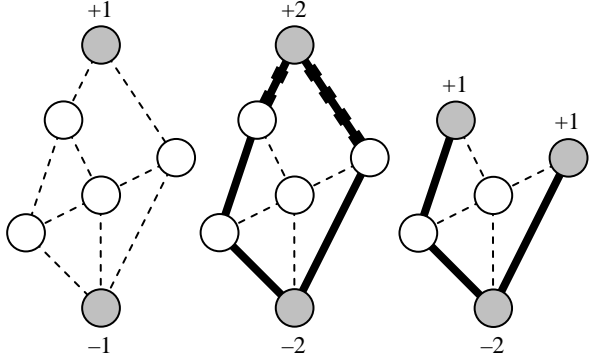

**Fig. 5.** Initial problem with one source and one sink node

**Fig. 6.** Maximize the flow, fix the new flow-out coefficients at the nodes and find the bottleneck links (layer 1, $F^1 = 2$)

**Fig. 7.** Remove the bottleneck links from the network and adjust the flow-out coefficients at the adjacent nodes

At each subsequent layer (Fig. 7 to Fig. 10) we have a bounded multi-source/multi-sink problem: a uniform flow from a set of sources to a set of sinks, where all rates of transmissions by sources and all rates of receptions by sinks increase proportionally in respect to each node's flow-out coefficient (either positive or negative). The multi-source/multi-sink problems arise since the LP problem at each successive layer is obtained by complete removal of the bottlenecks from the previous LP problem. By removing the bottlenecks we adjust correspondingly the flow-out coefficients of the adjacent nodes (to respect the flow conservation rule) and thus possibly produce new sources and sinks in the network. Except for the unicast problem of the first layer, the successive layer problems do not belong in general to the simple class of "network linear programs" [Fourer03].

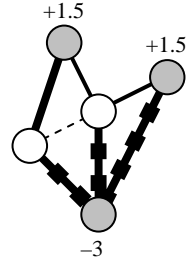 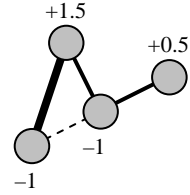 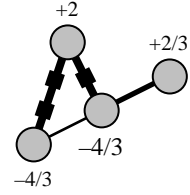

**Fig. 8.** Maximize the flow in the new sub-problem, fix the new flow-out coefficients at the nodes and find the new bottlenecks (layer 2, $F^2 = 1.5$)

**Fig. 9.** Again remove the bottleneck links from the network and adjust correspondingly the flow-out coefficients at the adjacent nodes

**Fig. 10.** Maximize the flow in the obtained new problem, fixing the new resulting flow-out coefficients at the nodes and find the new bottlenecks (layer 3, $F^3 = 4/3$)

We define the bounded multi-source/multi-sink problem at layer $l$ by the sets of nodes and links and by the flow-out coefficients for sources and sinks (all indexed with an upper index $l$) as follows:

- set of nodes $N^l$,
- set of links $(i, j) \in L^l$, where $i \in N^l$ and $j \in N^l$,
- flow-out coefficients $f_i^l$ for all $i \in N^l$
- at layer $l$ the max-flow solution yields the flow increase factor $F^l$ and the set of bottlenecks $B^l$, where $B^l \subset L^l$

Then, the equations for computing the sets $N^{l+1}$, $L^{l+1}$ and the flow-out coefficients $f^{l+1}$ of the next layer are as follows:

- $N^{l+1} = N^l$ (1)
- $L^{l+1} = L^l - B^l$ (2)
- $f_j^{l+1} = f_j^l \cdot F^l + \sum_{(i,j) \in B^l} (+1) + \sum_{(j,k) \in B^l} (-1)$ (3)

  add 1 for each incoming bottleneck link $(i, j)$ — subtract 1 for each outgoing bottleneck link $(j, k)$

After a certain number of applications of the max-flow objective with corresponding modifications of the problem, we will finally obtain a network having no source and sink nodes. At this point the iteration stops.

All links followed by the flow in the capillary routing are enclosed in bottlenecks of one of the layers.

In order to restore the original proportions of the flow, the flow increases, induced by the preceding max-flow solutions must all be compensated. The true value of flow $r_{i,j}$ traversing the bottleneck link $(i, j) \in B^l$ of layer $l$ is the initial single unit of flow divided by the product of the flow increase factors $F^i$ (where $1 \leq i \leq l$) of the present and all preceding layers:

$$r_{i,j} = \frac{1}{\prod_{i=1}^{l} F^i} \quad \text{where } l \text{ is the layer for which } (i, j) \in B^l \qquad (4)$$

The max-flow approach proves to be very stable, because it maintains all values of variables and parameters in the same order of magnitude (even for very deep layers with tiny loads) and also because it enables us to detect and correct errors in the flow-out coefficients of the LP problem generated for the next layer of capillary routing.

In the next subsection we show how to identify bottlenecks after the max-flow solution of the capillary routing layer is found.

### 2.3. Bottleneck hunting loop

In the example of Fig. 11 with three transmitting nodes and two receiving nodes, the flow can be proportionally increased at most by a factor of 4/3 and the bottleneck links are among four maximally loaded suspected links $\{a, b, d, e\}$, marked in Fig. 12 by thick dashes.

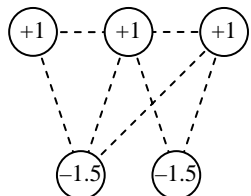

**Fig. 11.** An example of a bounded multi-source/multi-sink problem (obtained during construction of the capillary routing from a network with one source and one destination node)

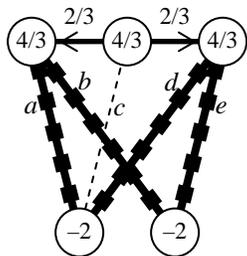

**Fig. 12.** A max-flow solution with the flow increase factor of 4/3, containing four maximally loaded candidate links $\{a, b, d, e\}$

At each layer, after minimizing the maximal load of links, the bottlenecks of the layer are discovered in a bottleneck hunting loop. At each iteration of the hunting loop, we minimize the load of the traffic over all links having maximal load and being suspected as bottlenecks. Links not maintaining their load at the maximum are removed from the suspect list. The bottleneck hunting loop stops if there are no more links to remove.

In the example of Fig. 12 the sum of loads of all four suspected links can be minimized (by an LP objective) to 3 (see Fig. 13). Now only three links $\{a, b, e\}$, marked by thick dashes, continue to maintain the maximal load. The sum of loads of three remaining suspected links can be further reduced to 2 (see Fig. 14). These two remaining links $\{b, e\}$, marked by thick dashes, maintained the maximal load at all times and are the true bottleneck links since the sum of their loads cannot be further reduced.

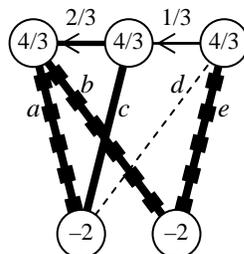 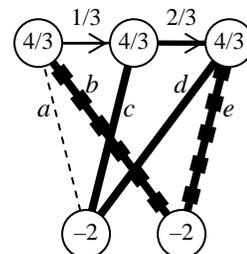

**Fig. 13.** Cost reduction applied to four fully loaded links of Fig. 12 reduces the load of suspected link $d$, and the suspect list is now $\{a, b, e\}$.

**Fig. 14.** Cost reduction applied to the three fully loaded links of Fig. 13 reduces the load of another suspected link $a$, and the true bottleneck links are $\{b, e\}$.

In this example the two bottlenecks are found in two iterations.

For capillary routing layers built simultaneously on 200 independent network samples each with 300 nodes (in average 2,555.7 links per network), Fig. 15 shows the decrease in the number of suspected links during the bottleneck hunting loop of each capillary routing layer from 1 to 10.

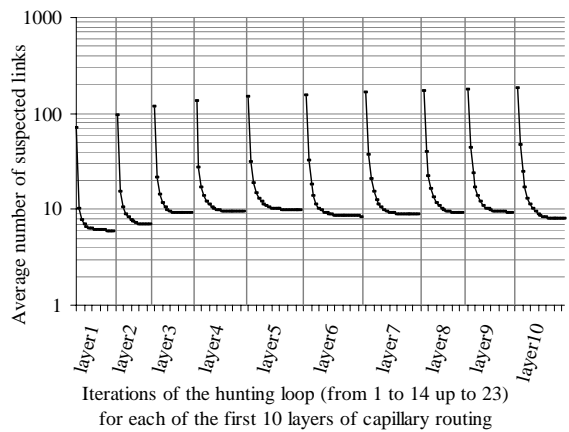

**Fig. 15.** Decrease of the number of suspected links during the bottleneck hunting loop of each of 10 capillary routing layers

At the end of each hunting loop (from 14 to 23 iterations) the suspect list consists of only true bottleneck links, in average between 5.9 and 9.9 bottlenecks per network.

## 3. Redundancy Overall Requirement (ROR)

The definition and equations of ROR are given in subsection 3.1. Computation of transmission FEC block size as a function of the packet loss rate *p* is presented in subsection 3.2. Equation of ROR for a particular case of very large FEC blocks is presented in subsection 3.3.

### 3.1. Definition of ROR

We assume a combination of a small static tolerance of the media stream to weak failures, with a dynamically added adaptive FEC for combating serious failures exceeding the tolerable packet loss rate.

For a given routing pattern, ROR is defined as the sum of all transmission rate overheads required from the sender for combating each non-simultaneous link failure in the route. For example, if the communication footprint consists of five links, and in response to each individual link failure the sender increases the packet transmission rate by 25%, then the ROR coefficient will be equal to the sum of these five FEC transmission rate increases, i.e. $ROR = 5 \cdot 25\% = 1.25$. If $P$ is the usual packet transmission rate and $P_l$ is the increased rate of the sender, responding to the failure of a link $l \in L$, where $L$ is the set of all links, then:

$$ROR = \sum_{l \in L}\left(\frac{P_l}{P} - 1\right) \quad (5)$$

Let us consider a long communication, and let *D* be the total failure time of a single network link during the whole duration of the communication. *D* is the product of the average duration of a single link failure, the frequency of a single link failure and the total communication time. According to equation (5):

$$D \cdot P \cdot ROR = D \cdot P \cdot \sum_{l \in L}\left(\frac{P_l}{P} - 1\right) \quad (6)$$
$$= \sum_{l \in L}(D \cdot P_l - D \cdot P) \quad (7)$$

Assuming one single link failure at a time and a uniform probability and duration of link failures, according to equation (7), $D \cdot P \cdot ROR$ is the number of adaptive redundant packets that the sender actually needs to transmit in order to compensate for all network failures occurring during the total communication time. Therefore ROR is a routing coefficient for computing the overall number of required redundant packets.

Redundant packets are injected into the original media stream for every block of *M* source packets. During streaming, *M* is supposed to stay constant. However, the number of redundant packets for each block of *M* media packets is variable, depending on the conditions of the erasure channel. The *M* source packets with their related redundant packets form a FEC block. By $FEC_p$ we denote the FEC block size chosen by the sender in response to a packet loss rate *p*. We assume that by default the media is streamed in FEC blocks of length of $FEC_t$ such that the flow has a static tolerance to weak losses $0 \leq t < 1$. When the loss rate *p* measured at the receiver is about to exceed the tolerable limit *t*, the sender increases its transmission rate by injecting additional redundant packets.

The random packet loss rate, observed at the receiver during the failure time of a link in the communication path, is the portion of the traffic still being routed toward the faulty link. Thus, a complete failure of a link *l* carrying a relative traffic load of $0 \leq r(l) \leq 1$ according to the routing pattern, produces at the receiver a packet loss rate equal to the same relative traffic load $r(l)$.

Equation (5) for ROR can thus be re-written as follows:

$$ROR = \sum_{l \in L \,|\, t \leq r(l) < 1}\left(\frac{FEC_{r(l)}}{FEC_t} - 1\right) \quad (8)$$

a sum over all links carrying a flow exceeding the tolerable loss limit

The links carrying the entire traffic are skipped in the sum index of equation (8), since the FEC required for the compensation of failures of such links is infinite. By construction (sections 2), none of the considered multi-path routing schemes pass their entire traffic through a non-critical single link.

### 3.2. Computing FEC block size

We compute the $FEC_p$ function assuming a Maximum Distance Separable (MDS) code [Seroussi86], [Schwarz02]. With an MDS code we can successfully decode the *M* source packets if we receive any *M* packets of the transmission FEC block.

In order to collect a mean of *M* packets at the receiver under random loss rate *p*, $M/(1-p)$ packets must be transmitted at the sender. However the probability of receiving $M-1$ packets or $M-2$ packets (which makes the decoding impossible) remains high. In order to maintain a very low probability $\delta$ of receiving less than *M* packets, we must send many more redundant packets in the block than is necessary to receive an average of *M* packets at the receiver side. We must fix the acceptable Decoding Error Rate (DER), such that $\delta \leq DER$, in order to compute the $FEC_p \geq M$ function.

The probability of having exactly $n$ losses (erasures) in a block of $N$ packets with a random loss probability $p$ is computed according to the binomial distribution:

$\binom{N}{n} \cdot p^n \cdot q^{N-n}$, where $\binom{N}{n} = \frac{N!}{n! \cdot (N-n)!}$ and $q = 1 - p$

The probability of having $N - M + 1$ or more losses, i.e. the decoding failure probability, is computed as follows:

$$\delta = \sum_{n=N-M+1}^{N} \binom{N}{n} \cdot p^n \cdot q^{N-n} \quad (9)$$

Therefore for computing the carrier block's minimal length for a satisfactory communication (i.e. $FEC_p$ function), it is sufficient to steadily increase the block length $N$ until the desired decoding error rate (DER) is met.

$FEC_p$ functions divided by $M$ (i.e. transmission rate increase factors $FEC_p / M$) are bounded above by $\log_p(DER)$ when $M = 1$ and below by $1/(1-p)$ when $M \to \infty$ (for packet loss rates much larger a very small DER). The higher the number of media packets in the block the closer the transmission rate increase can approach the lowest theoretical limit. For $M$ from 1 to 10 these transmission rate increase factors are plotted in Fig. 16 (for $DER = 10^{-5}$).

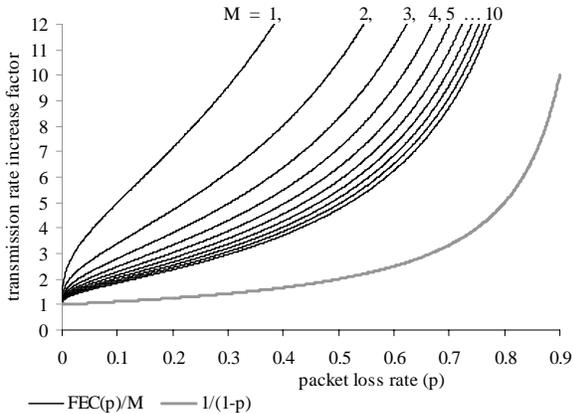

**Fig. 16.** Transmission rate increase factor as a function from the packet loss rate ($DER = 10^{-5}$)

### 3.3. Streaming with large FEC blocks

The larger the number of media packets $M$ in the FEC block, the smaller the cost of FEC overhead is, but the longer the buffering time at the receiver must be. For example VOIP with 20 ms sampling rate restricts the number of media packets $M$ in a single FEC block to 20 – 25 packets.

If the playback buffering time can be a couple of minutes long, with thousands of source packets in a FEC block (for example in packetized TV) we can assume that $FEC_p = M/(1-p)$. Although for large numbers of source packets MDS codes do not exist, other capacity-approaching LDPC [MacKay96], [Richardson01] or fountain codes [MacKay05] can decode a large block of source packets requiring only a very little excess of packets (in this context this excess can be ignored).

In such case, taking into account the above assumptions and equation (8), the ROR coefficient of a multi-path routing pattern is computed according to the following equation:

$$ROR = \sum_{l \in L \mid t \leq r(l) < 1} \left( \frac{1-t}{1-r(l)} - 1 \right) \quad (10)$$

Path diversity can be required in off-line large file downloads aiming at avoiding the idle times of the last kilometer bottleneck occurring due to arbitrary failures elsewhere, within the lossy Internet. Thanks to multi-path routing, the sender with an adaptive transmission rate can feed the last kilometer bottleneck link constantly at its maximal bandwidth (see [Nguyen02] and [Byers99] for video streaming from multiple servers). In this case also, the choice of the multi-path routing pattern can be rated by equation (10). Note that according to equations (8) and (10) the ROR coefficient of a routing pattern depends also on the static tolerance $t$ of the streaming media to weak failures.

## 4. Redundancy Overall Requirement in capillary routing

For capillary routing layers 1 to 10, we compute the average ROR coefficients simultaneously over several networks. The network samples are drawn from timeframes of a random walk MANET. Initially the nodes are randomly distributed on a rectangular area, and then, at every timeframe, they move according to a random walk algorithm. If two nodes are close enough (and are within the coverage range) then there is a link between them. At the same time we consider also streaming media at 15 different strengths of static FEC codes which tolerate small packet loss rates from 3.6% to 7.8% respectively (with an increment of 0.3%).

Fig. 17, represents a MANET with 115 nodes and 300 timeframes (each representing one network sample) divided into seven sets of network samples. For each set of samples and for each static FEC strength we plot the average ROR coefficient (over all considered network samples) as the routing layer increases. Fig. 17 shows that the overall requirement in adaptive FEC packets decreases with capillarization. The ROR coefficients of the routing samples are computed assuming a short playback buffering time

according to equation (8), where the FEC block size (as function of the packet loss rate $p$) is computed according to equation (9), the number of media packets ($M$) per transmission block is 20 and the desired decoding failure rate (DER) is $10^{-5}$.

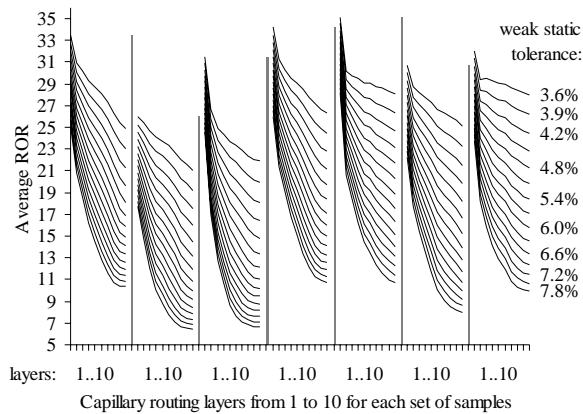

**Fig. 17.** Average ROR as a function from the capillary routing layer

Fig. 18 represents a MANET with 120 nodes and 150 timeframes divided into four sets of network samples. The upper 15 curves similarly to the curves of Fig. 17 are computed according to equations (8) and (9), where $M = 20$ and $DER = 10^{-5}$. However, the lower 15 curves of Fig. 18 are computed according to equation (10) for streaming with large FEC blocks.

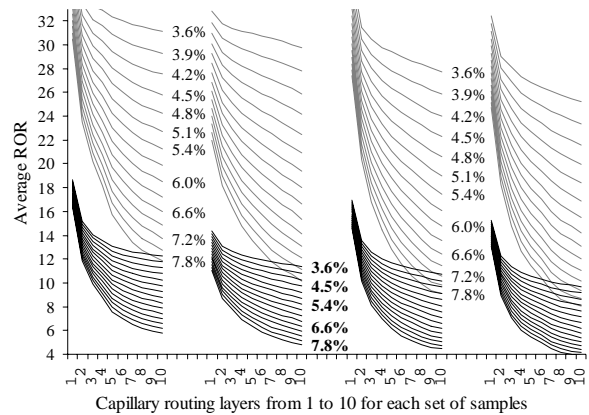

**Fig. 18.** Average ROR computed assuming real-time streaming (the group of curves above) and off-line streaming (the group below)

When streaming with large blocks the Redundancy Overall Requirement is twice as low as in streaming with restricted playback buffering time, but the capillarization of routing is beneficiary in both cases.

Logically, the ROR curve of the media stream is shifted down as the statically added tolerance increases, but the increase of the weak static tolerance emphasizes the efficiency gain achieved by capillarization. The drawback of path diversity in general is that by forming long paths we increase the number of links in the communication footprint raising the overall failure rate and thus possibly increasing the overall requirement in FEC codes. However, Fig. 17 and Fig. 18 show that despite the communication footprint becomes larger; with the routing patters built by the capillary routing algorithm the requirement in redundant packets decreases noticeably most of the time.

## 5. Conclusion

The reliability issues of packetized real-time streaming are of growing importance. Commercial real-time streaming applications however do not consider channel coding at the packet level as a serious solution for improving the reliability of communication. That is because in single path communications, even heavy FEC overheads cannot protect against failures lasting more than the short duration of the playback buffer. Recent studies demonstrated that path diversity makes FEC applicable for real-time streaming. By studying a wide range of routing topologies, we show that combination of channel coding with appropriate multi-path routing allows reliable real-time streaming with a low overall requirement in FEC codes.

For this purpose we introduced a layer by layer strategy for building multi-path capillary routing patterns. The first layer provides a simple multi-path solution. As the layer number increases, the underlying routing pattern relies on the network more securely. Unlike max-flow or shortest path solutions, for a given source and destination, by construction (section 2) there exists only one solution of capillary routing.

We introduced ROR coefficient, a method for rating multi-path routing patterns by a single scalar value. The ROR rating corresponds to the total redundancy overhead that the sending node must provide in order to combat the losses occurring from non-simultaneous failures of links in the communication path. Despite the fact that the spreading out of the routing results in the increase of the overall failure rate of underlying links, with capillarization the overall requirement in adaptive FEC packets decreases substantially.

Capillary routing can be applicable to multi-hop mobile wireless networks, where wireless content can be streamed to and from the user via multiple base stations; or to the public internet, where, if the physical routing cannot be accessed, an overlay network can be used [Guven04]. We hope that our investigation will provide some guidelines for future design of path diversity-based real-time streaming systems.